\documentclass[reprint,
superscriptaddress,
frontmatterverbose, 
preprintnumbers,
nofootinbib,
nobibnotes,
amsmath,amssymb,
aps,
pra,
floatfix,
]{revtex4-1}

\usepackage{graphicx}
\usepackage{dcolumn}
\usepackage{bm}
\usepackage{hyperref}
\usepackage[mathlines]{lineno}

\begin{document}


\title{Focused linearly-polarized light scattering from a silver nanowire: Experimental characterization of optical spin-Hall effect}

\author{Diptabrata \surname{Paul}}
\affiliation{Department of Physics, Indian Institute of Science Education and Research (IISER), Pune 411008, India}
\author{Deepak K. \surname{Sharma}}
\affiliation{Department of Physics, Indian Institute of Science Education and Research (IISER), Pune 411008, India}
\affiliation{Present address: Laboratoire Interdisciplinaire Carnot de Bourgogne, UMR 6303 CNRS, Université de Bourgogne Franche-Comté, 9 Avenue Alain Savary, 21000 Dijon, France}
\author{G V Pavan \surname{Kumar}}
\email{pavan@iiserpune.ac.in}
\affiliation{Department of Physics, Indian Institute of Science Education and Research (IISER), Pune 411008, India}
\affiliation{Center for Energy Science, Indian Institute of Science Education and Research (IISER), Pune 411008, India}


\begin{abstract}

Spin-orbit interactions (SOI) are a set of sub-wavelength optical phenomenon in which spin and spatial degrees of freedom of light are intrinsically coupled. One of the unique example of SOI, spin-Hall effect of light (SHEL) has been an area of extensive research with potential applications in spin controlled photonic devices as well as emerging fields of spinoptics and spintronics. Here, we report our experimental study on SHEL due to forward scattering of focused linearly polarized Gaussian and Hermite-Gaussian ($\textrm{HG}_{10}$) beams from a silver nanowire (AgNW). Spin dependent anti-symmetric intensity patterns are obtained when the polarization of the scattered light is analysed. The corresponding spin-Hall signal is obtained by computing the far-field longitudinal spin density ($s_3$). Furthermore, by comparing the $s_3$ distributions, significant enhancement of the spin-Hall signal is found for $\textrm{HG}_{10}$ beam compared to Gaussian beam. The investigation of the optical fields at the focal plane of the objective lens reveals the generation of longitudinally spinning fields as the primary reason for the effects. The experimental results are corroborated by 3-dimensional numerical simulations. The results lead to better understanding of SOI and can have  direct implications on chip-scale spin assisted photonic devices.

\end{abstract}

\maketitle

\section{\label{sec:level1}Introduction}

Along with energy and linear momentum, angular momentum (AM) represents the most important dynamical parameters of light \cite{PhysRevA.45.8185, doi:10.1002/lpor.200810007, Bekshaev_2011}. The AM of light can be decomposed into two parts, spin angular momentum (SAM), which is related to the circular polarization of a light beam and orbital angular momentum (OAM), related to the helical phase front of a beam \cite{PhysRevA.45.8185, Yao:11, Bekshaev_2011,Molina-Terriza2007}. In recent times, these concepts have been exploited in context of spin-orbit interaction (SOI), an optical phenomenon in which polarization and position of light are intrinsically coupled \cite{PhysRevA.46.5199, bliokh2015spin, Aiello2015}.

One of the unique examples of SOI is spin-Hall effect of light (SHEL) \cite{PhysRevLett.93.083901}, where equivalent to its electronic counterpart \cite{PhysRevLett.83.1834}, a transverse shift in the scattered beam location is perceived due to the transverse spin flow of the impinging light beam on a dielectric interface \cite{hosten2008observation,bliokh2013goos,PhysRevA.85.023842,goswami2014simultaneous} or a scatterer \cite{PhysRevLett.102.123903, sharma2018spin}. Further, the transverse shift in the scattered beam is opposite for opposite spins of the impinging beam, allowing us to distinguish between them \cite{sharma2018spin}. But, since SOI effects are weak for paraxial beams, observation of transverse shift in scattering of linearly polarized beams becomes difficult due to absence of net transverse spin flow. The situation changes drastically in case of non-paraxial light beams, where depending on the numerical aperture (NA), higher AM can be generated \cite{bliokh2015spin, PhysRevA.82.063825}, thus leading to enhanced SOI \cite{doi:10.1063/1.2402909, nieminen2008angular}. To this end, strong focusing of light has been an area of extensive research in recent years with implications in studies such as optical manipulation \cite{zhao2009direct, doi:10.1021/jz401381e}, sub-wavelength position sensing \cite{PhysRevLett.104.253601, neugebauer2018weak}, AM inter-conversion \cite{PhysRevLett.99.073901, PhysRevB.99.075155}. In this context, the interaction of the optical fields at the focus with a nanoscopic object and the resultant SOI at sub-wavelength scale can have potential utilization in various photonic applications such as generation of structured fields, controlling optical wave propagation, optical nano-probing and refractive index sensor \cite{PhysRevB.88.121410, o2014spin,cardano2015spin, PhysRevLett.104.253601, neugebauer2018weak, zhou2018photonic}.

\begin{figure}[b]
\includegraphics{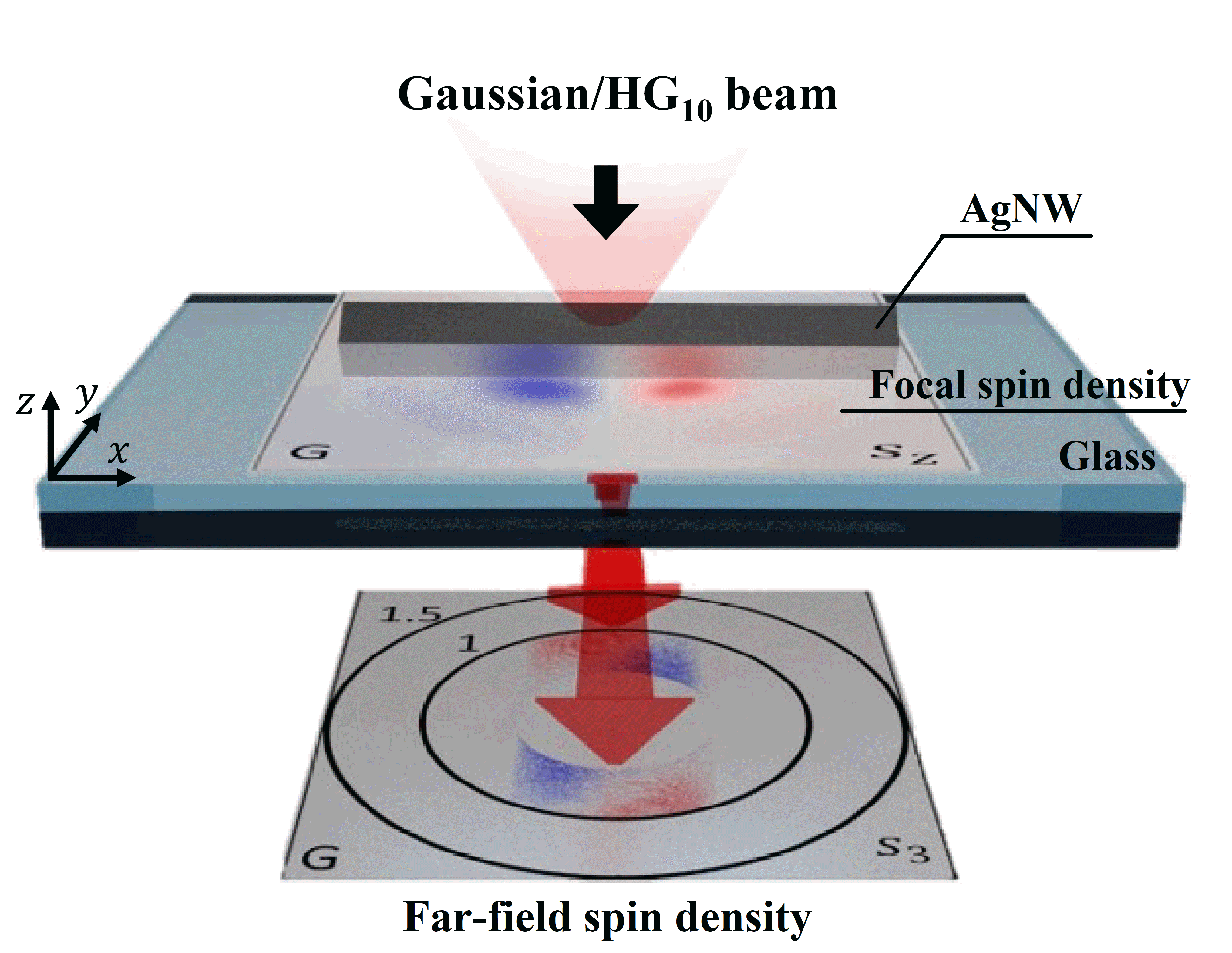}
\caption{\label{fig:1} (color online) Schematic of the measurement scheme of SHEL. The focal longitudinal spin density ($s_z$) is scattered by an AgNW and can be observed in the Fourier plane image as far-field longitudinal spin distribution ($s_3$).}
\end{figure}

Motivated by this, we study the longitudinal spin density through optical interaction between a quasi one dimensional scatterer, namely a single crystalline silver nanowire (AgNW) on a glass substrate and focused linearly polarized Gaussian and $\textrm{HG}_{10}$ beams as shown in Fig. \ref{fig:1}. Polarization analysis of the forward scattered light in the Fourier plane (FP) reveals anti-symmetric distribution of intensity for opposite circular polarization states with respect to long axis of the AgNW. The pattern inverts for left circular polarization (LCP) analyzed case with respect to that of right circular polarization (RCP) analyzed one, revealing intrinsic SHEL \cite{sharma2018spin}. The difference between the two FP intensity distributions quantify the spin-Hall signal as well as far-field longitudinal spin density ($s_3$). Furthermore, the comparison of $s_3$ for Gaussian beam with respect to that of $\textrm{HG}_{10}$ beam reveals higher spin-Hall signal for $\textrm{HG}_{10}$ beam. The enhancement factor of longitudinal spin distribution for $\textrm{HG}_{10}$ beam with respect to that of Gaussian beam is further quantified from experimentally measured and numerically simulated data.

In the following sections, we describe the nature of the optical fields at the focus and show the existence of longitudinally spinning fields at focal plane. Furthermore, by analyzing the scattering pattern of the focal optical fields from a AgNW, emerging SHEL is observed and resulting far-field longitudinal spin density is measured.

\section{\label{sec:level2}Theoretical Description}

\subsection{Description of focal optical fields}

Focusing of paraxial optical beams through an objective lens leads to complex distribution of optical fields at the focal plane. The nature of polarized optical fields at the focus of an objective lens can be theoretically calculated using Debye-Wolf integral, originally proposed by Richards and Wolf \cite{doi:10.1098/rspa.1959.0199, doi:10.1098/rspa.1959.0200}. When an paraxial light beam with electric field $\mathbf{E_{in}}$ and wave vector $\mathbf{k}=k\mathbf{\hat{z}}$ passing through medium of refractive index $n_1$ is incident on an objective lens of numerical aperture, $\textrm{NA} = n_1 \sin(\theta_{\mathrm{max}})$ and having focal length $f$, then the complex field distribution at the focal plane can be obtained by,

\begin{subequations}
\label{eq:1}
\begin{eqnarray}
\mathbf{E}(\rho,\varphi,z) =&& -\frac{\mathrm{i} kf\mathrm{e}^{-\mathrm{i}kf}}{2\pi}\int\displaylimits_{0}^{\theta_{\mathrm{max}}}\int\displaylimits_{0}^{2\pi}\mathbf{E_{ref}(\theta,\phi)}\mathrm{e}^{\mathrm{i} kz\cos\theta}\nonumber\\
&&\mathrm{e}^{\mathrm{i} k\rho\sin\theta\cos(\phi-\varphi)}\sin\theta \,\mathrm{d}\phi\, \mathrm{d}\theta
\end{eqnarray}

\begin{eqnarray}
\mathbf{E_{ref}} =&& [t^\mathrm{s}[\mathbf{E_{in}}\cdot\mathbf{n_\phi}]\mathbf{n_\phi}+t^\mathrm{p}[\mathbf{E_{in}}\cdot\mathbf{n_\rho}]\mathbf{n_\theta}]\sqrt{\cos\theta}
\end{eqnarray}
\end{subequations}

\begin{figure}[t]
\includegraphics{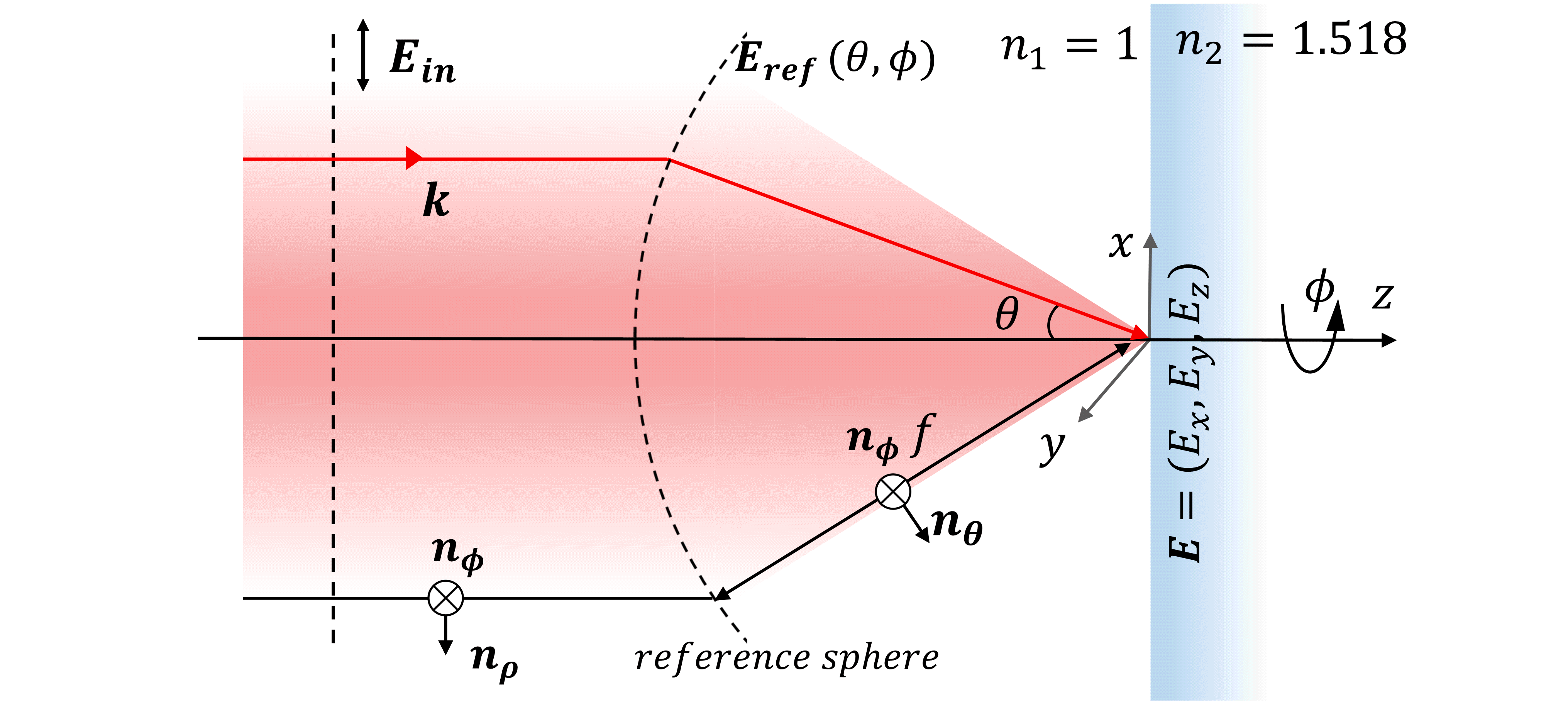}
\caption{\label{fig:2} (color online) Schematic of the focusing configuration. The incident field is denoted by $\mathbf{E_{in}}$ and arrow indicate the incident polarization. Spherical lens surface is shown by the reference sphere arc with focal length $f$. The electric field after refraction at the lens surface is given by $\mathbf{E_{ref}}$. $\mathbf{E} = (E_x,E_y,E_z)$ represent the focal electric field components.}
\end{figure}

The geometric representation of the system is given in Fig. \ref{fig:2} and closely follows ref. \cite{novotny_hecht_2006}. The formulation takes into account both the refraction of the paraxial beam at the spherical lens surface (reference sphere) given by $\mathbf{E_{ref}}(\theta,\phi)$ and the non-paraxial effects there after. $t^\mathrm{s}$ and $t^\mathrm{p}$ are the transmission coefficients corresponding to $\mathbf{s}$ and $\mathbf{p}$ polarized electric field components respectively. $\mathbf{n_\rho}$ and $\mathbf{n_\phi}$ represents the unit vectors of a cylindrical coordinate system whereas the unit vectors of a spherical polar coordinate are given by $\mathbf{n_\theta}$ and $\mathbf{n_\phi}$, origin of the coordinate system being the focal point $(x,y,z) = (0,0,0)$, as shown in Fig. \ref{fig:2}. $n_1$ and $n_2$ represent the medium refractive index.

\begin{figure}[t]
\includegraphics{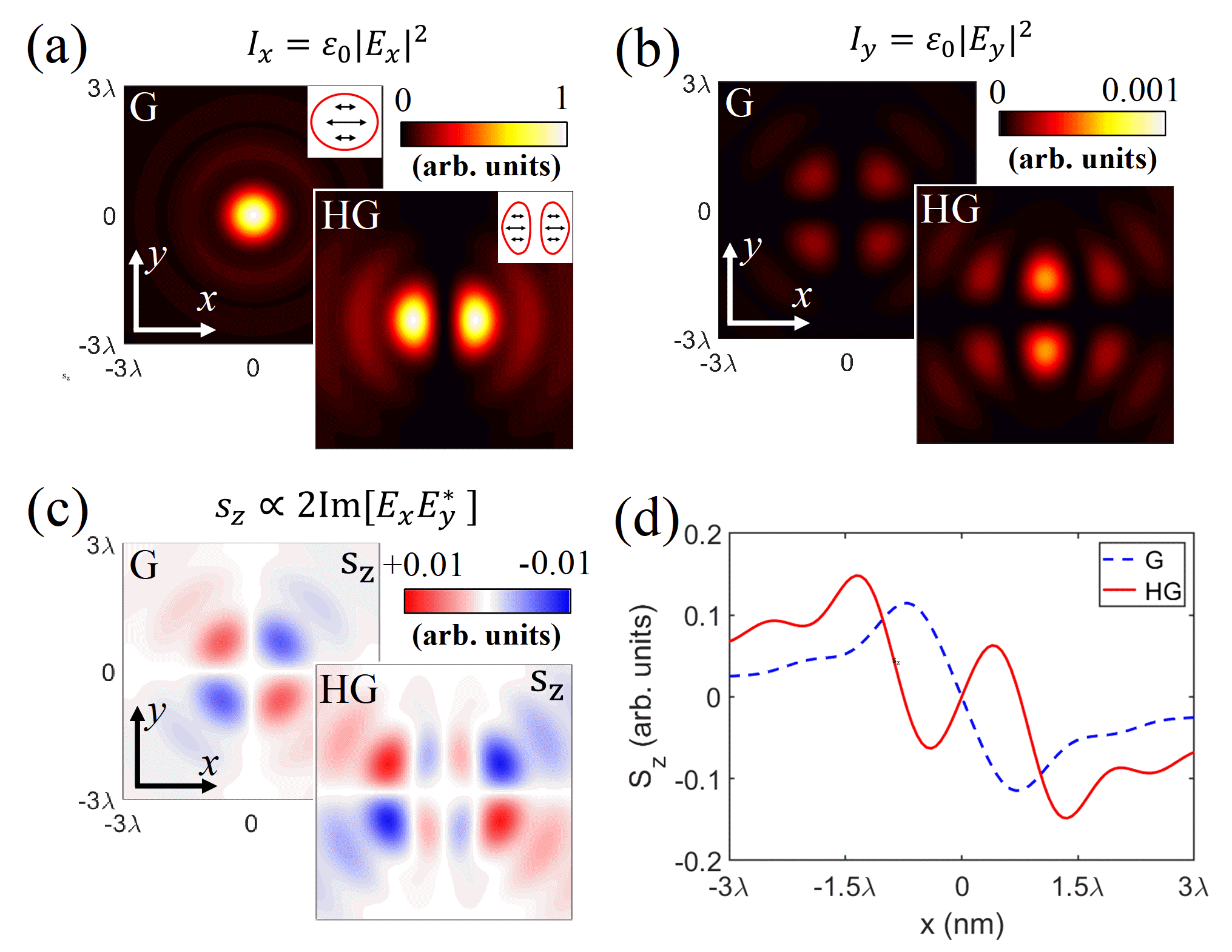}
\caption{\label{fig:3} (color online)  Theoretically calculated focal intensity profiles corresponding to (a) $x$ polarized field component, (b) $y$ polarized field component of the total optical field for Gaussian and $\mathrm{HG_{10}}$ beams. Insets in (a) shows the $x$ polarized paraxial beam profiles. (c) represents longitudinal spin density ($s_z$) at the focus for Gaussian and $\mathrm{HG_{10}}$ beams. (d) Comparative plot of longitudinal spin density  distribution ($S_z$) for Gaussian (dashed blue line) and $\mathrm{HG_{10}}$ beam (solid red line).}
\end{figure}

Using Eq. \ref{eq:1}(a), the electric field profiles, $\textbf{E} = (E_x,E_y,E_z)$, at the focal plane of a 0.5 NA lens is calculated for linearly polarized ($x$ polarized) paraxial Gaussian(G) and Hermite-Gaussian beam of order $m=1$, $n=0$ ($\mathrm{HG_{10}}$) at wavelength $\lambda=633$ nm, propagating along $z$ axis. Here $m$ and $n$ represent number of nodal lines along $x$ and $y$ axis respectively. The intensity distribution corresponding to the $x$ polarized  ($I_x=\varepsilon_{0}{|E_{x}|}^2$) and $y$ polarized ($I_y=\varepsilon_{0}{|E_{y}|}^2$) component of the total focal optical field for Gaussian(G) and $\mathrm{HG_{10}}$ (HG) beams are given in Figs. \ref{fig:3}(a) and \ref{fig:3}(b) respectively. The intensity distribution values are normalized with respect to maximum value of total intensity, $I=\varepsilon_{0}{|E|}^2$. The paraxial polarization profile of the beams are given in \ref{fig:3}(a) insets. Close inspection of the focal intensity profile in Fig. \ref{fig:3}(b) reveal that compared to $I_y$ of Gaussian beam, $I_y$ of $\mathrm{HG_{10}}$ beam is significantly higher, although their distributions differ. The ratio of maximum value of $I_y$ of $\mathrm{HG_{10}}$ beam to that of Gaussian beam turns out to be $\approx 2.97$.

\subsection{Spin density of optical fields}

One of the unique features that emerge due to focusing of linearly polarized paraxial beam is presence of spinning electric fields. This can be quantified using spin density \cite{bliokh2015transverse}, given by,
\begin{equation}
\mathbf{s}\propto \textrm{Im}\left[\varepsilon_0 (\mathbf{E^*} \times \mathbf{E} ) + \mu_0 (\mathbf{H^*} \times \mathbf{H} )\right].
\label{eq:2}
\end{equation}
The symmetric nature of Eq. (\ref{eq:2})  allows us to consider the contribution of either the electric field (\textbf{E}) part or the magnetic field (\textbf{H}) part. Since a plasmonic scatterer responds more towards the electric field part of the incident field, we consider the electric field contributions only. The spinning fields represented by Eq. (\ref{eq:2}) can be decomposed into three components which can be either longitudinal or transverse with respect to the propagation axis ($z$ axis) \cite{Aiello_2016}. The longitudinally spinning optical fields, given by:
\begin{equation}
s_z\propto2\,\textrm{Im}[E_x E_y^*]\propto s_3.  
\label{eq:3}
\end{equation}
can be quantified in the far-field through measuring the Stokes parameter $s_3$, which determines the degree of circular polarization of a planar optical field \cite{PhysRevE.66.016615}.

Theoretically calculated distribution of $s_z$ at the focus of $x$ polarized Gaussian and $\mathrm{HG_{10}}$ beams are given in Fig. \ref{fig:3}(c). For simplicity, we have ignored the proportionality constants in the equations. The values have been normalized with respect to maximum of $I$. A quantitative measure of the $s_z$ at the focus can be obtained by computing longitudinal spin density distribution ($S_z$) of one half of the focal plane, i.e. by integrating over the one of the spatial coordinates $y = 0$ to $3\lambda$:
\begin{equation}
S_z(x) = \int\displaylimits_{y=0}^{3\lambda} s_z(x,y) dy.
\label{eq:4}
\end{equation}

Comparing the $s_z$ in Fig. \ref{fig:3}(c) for Gaussian and $\mathrm{HG_{10}}$ beams, it is evident that the $x$ polarized $\mathrm{HG_{10}}$ beam possesses significantly higher $s_z$ with respect to that of $x$ polarized Gaussian beam. The same is reflected in Fig. \ref{fig:3}(d), which show the comparative magnitude of $S_z$ for $x$ polarized Gaussian (dashed blue line) and $\mathrm{HG_{10}}$ beams (solid red line) as a function of $x$ coordinates at the focal plane. The enhancement of $s_z$ for $\mathrm{HG_{10}}$  beam with respect to that of Gaussian beam can be obtained by calculating the ratio of extremum value of $S_z$, which in this case is $\eta_{theory}= 1.29$. The enhancement factor, $\eta_{theory}$ is dependent on the NA of the objective lens.

\section{\label{sec:level3}Experimental Implementation}

\begin{figure}[b]
\includegraphics{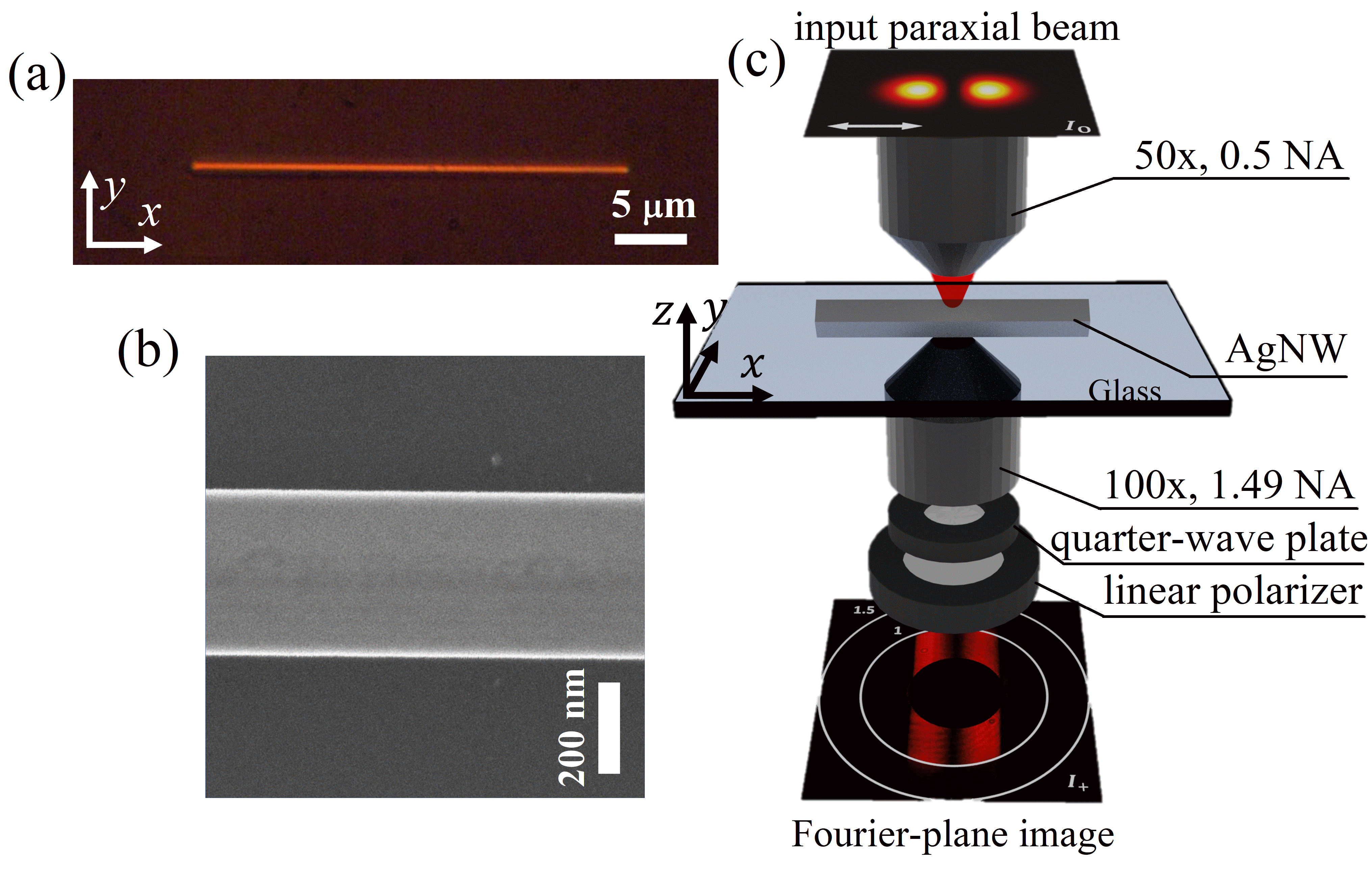}
\caption{\label{fig:4} (color online) (a) Optical bright-field image of the AgNW used for our experiments. (b) Magnified FESEM image of top view of an AgNW section. The schematic of experimental setup for the elastic forward scattering is given in (c).}
\end{figure}

The experimental implementation in observing the SHEL as well as measuring the longitudinal spin density relies on elastic scattering of the focal optical fields. Various probe geometries can be used for studying such light matter interactions such as, spherical particles \cite{PhysRevLett.114.063901, PhysRevX.8.021042, yang2018optical, PhysRevA.97.043823}, gold chiral geometry \cite{wozniak2019interaction}. We use a pentagonal-cross sectional silver nanowire (AgNW) for our study. Chemically synthesized single crystalline AgNWs of diameter $\sim 350$ nm were drop casted onto a glass substrate \cite{doi:10.1021/nl034312m}. Fig. \ref{fig:4}(a) shows an optical image of the AgNW used for our experiments. Fig. \ref{fig:4}(b) shows scanning electron micrograph of a nanowire section. The AgNW is illuminated by Gaussian beam and $\mathrm{HG_{10}}$ beam, prepared by projecting Gaussian beam onto a spatial light modulator with blazed hologram, at wavelength $\lambda=632.8$ nm. The paraxial (Gaussian/$\mathrm{HG_{10}}$) beam, with polarization parallel to the long axis of the AgNW ($x$ axis), is focused at the center of the AgNW placed at ($x=0$, $y=0$), with a $0.5$ NA objective lens.

The forward scattered light is collected using an oil immersion objective lens having 1.49 NA and then analyzed using a combination of linear polarizer and quarter wave plate. The schematic of forward scattering setup is depicted in Fig. \ref{fig:4}(c). The scattered light in the Fourier plane (FP) is captured by relaying the back-focal plane of the collection objective lens onto a CCD \cite{kurvits2015comparative, doi:10.1002/3527600434.eap817}. Detailed experimental setup used for the study is similar to our previous reports and is given in Appendix \ref{sec:ap3} \cite{sharma2018spin, doi:10.1021/acsphotonics.8b01220}. The use of low NA (0.5 NA) excitation and high NA (1.49 NA) collection objective lens facilitates the collection of scattered light in higher angles in the FP. The un-scattered light, which dominates the incident NA part of the FP intensity distribution, is omitted for analysis. 

\section{\label{sec:level4}Results and Discussion}

\subsection{Nanowire as a scatterer}

An AgNW placed at the focal plane of an objective lens mimics the effect of a nanoscopic strip diffraction \cite{Bekshaev_2017,bekshaev2020optical}. Scalar diffraction theory and Babinet’s principle can be applied to obtain the diffraction pattern in case of strip diffraction of paraxial beams. But due to the complex and vectorial nature of the non-paraxial optical fields near to the focal plane, full scattering theory \cite{born_wolf_bhatia_clemmow_gabor_stokes_taylor_wayman_wilcock_1999} has to be considered for nanowire diffraction. To this end, the scattered optical field intensity distribution in FP from AgNW has been previously used for probing the SOI with the incident optical fields \cite{sharma2018spin, doi:10.1021/acsphotonics.8b01220}.

The scattered optical field from the AgNW can be collected in two distinct regions in the far-field, depending on its origin. These are sub-critical or allowed region ($\textrm{NA}<1$) and super-critical or forbidden region ($\textrm{NA}>1$) in FP. The longitudinally polarized field ($E_x$) component of the total optical field gets maximally scattered in the sub-critical region. The transversely polarized optical field ($E_y$) component leads to near-field (NF) accumulation at the AgNW edges \cite{doi:10.1021/nn201648d}. The evanescent NF gets partially converted into propagating waves at the air-glass interface and can be observed in the super-critical region of the far-field \cite{novotny_hecht_2006}. Hence, elastic scattering from an AgNW allows us to route a portion of the scattered light either in the sub-critical region or in the super-critical region by engineering the incident linear polarization state as longitudinal or transverse to the wire. To this end, in our previous report, the resultant SOI effects due to forward scattering of incident circularly polarized light beams in similar experimental configuration have been reported to be prominent within the sub-critical region of FP \cite{sharma2018spin}. Hence, by analyzing of the FP intensity distribution in the sub-critical region due to scattering of longitudinally polarized paraxial beams, we can probe the resultant SOI effects as well as extract the characteristics of the focal optical field.

\begin{figure*}[t]
\includegraphics{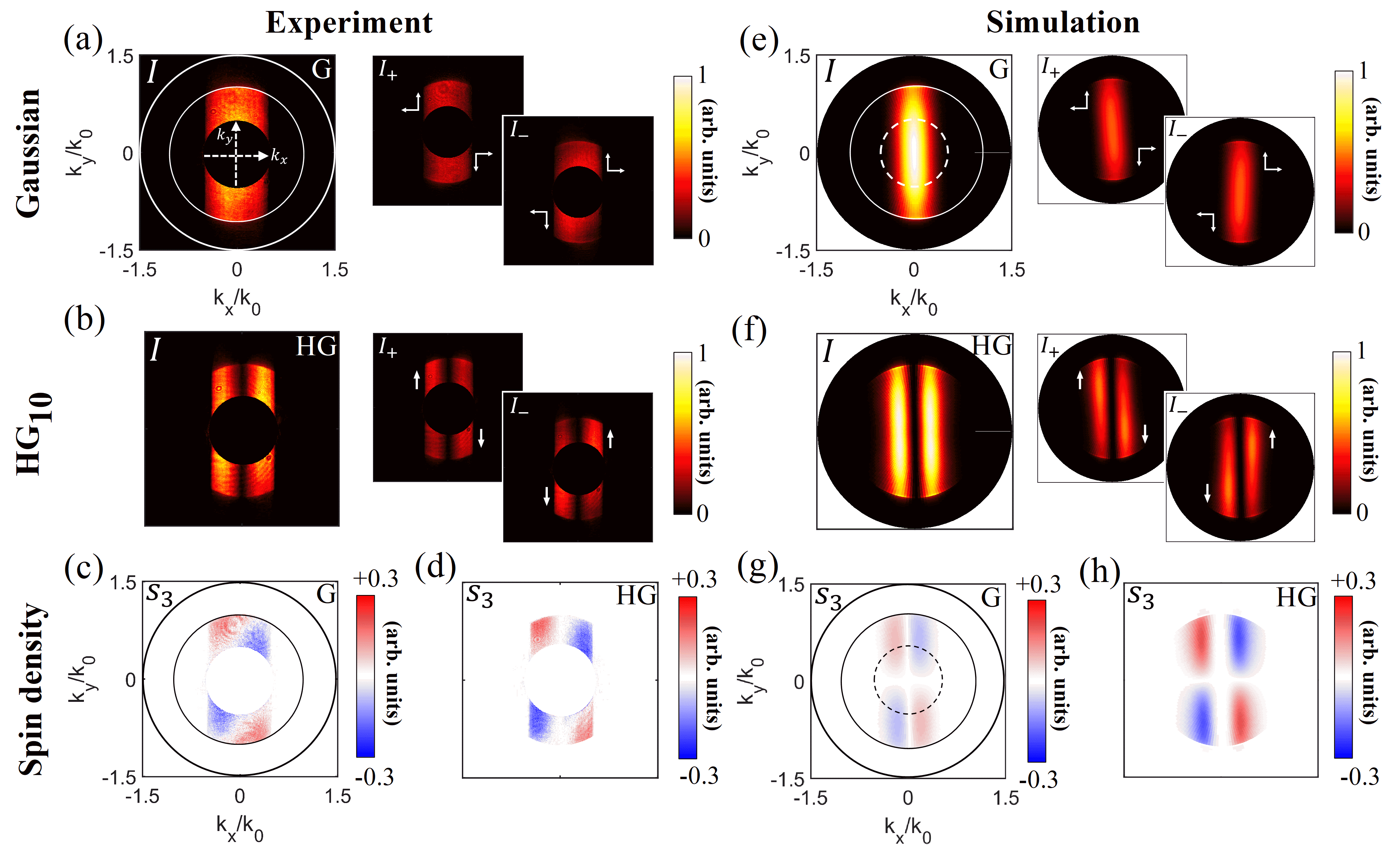}
\caption{\label{fig:5} (color online) Experimentally measured far-field intensity patterns for (a) Gaussian beam, (b) $\textrm{HG}_{10}$ beam. $I$ represents total intensity, $I_+$ and $I_-$ represents LCP and RCP analyzed FP intensity distribution respectively. Inner and outer circles in (a) represent the critical angle at air-glass interface and the collection limit of the objective lens NA respectively. White arrows in $I_+$ and $I_-$ indicate the bias of the intensity lobe(s). Black disk at the center of the FP indicate the blocked incident NA region. Far-field longitudinal spin density ($s_3$) patterns are shown in (c) Gaussian beam and (d) $\textrm{HG}_{10}$ beam. Numerically simulated far-field intensities $I$, $I_+$, $I_-$ for Gaussian and $\textrm{HG}_{10}$ beams are given in (e) and (f) respectively. (g) and (h) represent numerically calculated $s_3$ for Gaussian and $\textrm{HG}_{10}$ beams respectively. The inner dashed white circle in (e) and dashed black circle in (g) represent the incident NA region equivalent to experimental counterpart.}
\end{figure*}

\subsection{Experimentally measured scattering intensity}

First, we examine the elastic scattering of optical fields due to focusing of longitudinally polarized paraxial Gaussian and $\textrm{HG}_{10}$ beams from an AgNW. Experimentally measured FP intensities of forward scattered Gaussian and $\textrm{HG}_{10}$ beams are shown in Figs. \ref{fig:5}(a) and \ref{fig:5}(b) respectively. The inner and outer white circles in Fig. \ref{fig:5}(a) indicate the critical angle at air-glass interface and collection limit of the objective lens respectively. $I$ represents the un-analyzed scattered FP intensity distribution. $I_{+}$ and $I_{-}$ denotes the left circular polarization (LCP) and right circular polarization (RCP) analyzed FP intensity distribution respectively. Absence of light in the super-critical region ($\textrm{NA} = k_x/k_0=k_y/k_0>1$) in the measured FP intensity distribution indicates the minimal accumulation of NF at the NW edges. The incident NA is blocked and is shown by a black disk at the center. 

The FP intensity distribution corresponding to $I$ of Gaussian beam in Fig. \ref{fig:5}(a) has only one lobe where as the scattering intensity pattern due to $\textrm{HG}_{10}$ beam in Fig. \ref{fig:5}(b) has two intensity lobes due to the presence of nodal line in its incident beam intensity profile (see Fig. \ref{fig:3}(a)). Both the intensity patterns corresponding to $I$ in Figs. \ref{fig:5}(a) and \ref{fig:5}(b) show symmetric pattern with respect to $k_x/k_0$ axis (parallel to AgNW long axis). In contrast, the FP intensity distribution corresponding to $I_+$ and $I_-$ of Gaussian beam exhibit shift along $k_x/k_0<0$ and $k_x/k_0>0$ respectively for $k_y/k_0>0$ region (vice versa for $k_y/k_0<0$ region), revealing anti-symmetric scattering pattern, as depicted by the white arrows in Fig. \ref{fig:5}(a) (see Appendix \ref{sec:ap2}). For $\textrm{HG}_{10}$ beam, the analysed $I_+$ and $I_-$ component exhibit shift along $k_y/k_0<0$ and $k_y/k_0>0$ direction for $k_x/k_0>0$ region respectively (and vice versa for $k_x/k_0<0$ region), as depicted by the white arrows in Fig. \ref{fig:5}(b). The opposite shifts of the intensity distribution of $I_+$ with respect to that of $I_-$ indicates the presence of partial circular polarization in the optical fields \cite{Neugebauereaav7588}. This spin dependent opposite angular shift of the lobe(s) in the FP corresponding to $I_+$ and $I_-$ is analogous to SHEL \cite{PhysRevLett.93.083901,Neugebauereaav7588,sharma2018spin}. Thus, the magnitude of this spin-Hall signal and the consequent presence of the spin in the FP can be obtained by calculating far-field longitudinal spin density ($s_3$) as, $s_3\propto I_{+} - I_{-}$. Experimentally measured $s_3$ for forward scattered Gaussian and $\textrm{HG}_{10}$ beams are shown in Figs. \ref{fig:5}(c) and \ref{fig:5}(d) respectively. The experimentally measured intensity values are normalized with respect to the maximum value of $I$. 

\subsection{Numerically simulated scattering intensity}

The experimental results are corroborated by full wave 3-dimensional finite element method (FEM) simulation. A geometry having pentagonal cross-section with diameter $350$ nm and length $5$ $\mathrm{\mu m}$ is used for modelling of the NW. Material of the NW is mimicked by matching the refractive index with that of Ag at $633$ nm wavelength \cite{johnson1972optical}. Meshing of the system is done by free tetrahedral geometry of size $35$ nm. The NW is illuminated with excitation beams (Gaussian, $\textrm{HG}_{10}$) at $\lambda=633$ nm. The beam waist at the air-glass interface is fixed to $633$ nm to mimic similar focusing conditions as in our experiments. Reciprocity arguments are used for near field to far-field transformation of the scattered optical field \cite{yang2016near}.

Numerically simulated FP intensity distribution of forward scattered light for Gaussian and $\textrm{HG}_{10}$ beams are shown in Figs. \ref{fig:5}(e) and \ref{fig:5}(f) respectively. $I$ represents the total scattered intensity. $I_+$ and $I_-$ represent the LCP and RCP analyzed FP intensities respectively. Similar to Fig. \ref{fig:5}(a), the numerically calculated intensity distribution for $I_+$ and $I_-$ of Gaussian beam in Fig. \ref{fig:5}(e) have opposite shifts of the intensity lobe(s) along $k_x/k_0$ axis, hence exhibiting anti-symmetric scattering pattern, as shown by the white arrows (see Appendix \ref{sec:ap2}). The numerically simulated FP intensity distribution corresponding to $I_+$ and $I_-$ of $\textrm{HG}_{10}$ beam in Fig. \ref{fig:5}(f), similar to its experimental counterpart in Fig. \ref{fig:5}(b), exhibit shift along $k_y/k_0<0$ and $k_y/k_0>0$ direction for $k_x/k_0>0$ region respectively (and vice versa for $k_x/k_0<0$ region), indicated by the white arrows. The opposite shifts for $I_+$ and $I_-$ components are analogous to angular spin-Hall shift. Corresponding to this angular shift, the $I_+$ and $I_-$ components of the the scattered field at the  $x$-$y$ plane ($z=0$ plane, air-glass interface plane) also exhibit opposite positional shift for Gaussian and $\textrm{HG}_{10}$ beams (see Appendix \ref{sec:ap1}). Finally, the spin-Hall signal can be is obtained by computing the far-field longitudinal spin density ($s_3$) as, $s_3\propto I_{+} - I_{-}$. Numerically simulated $s_3$ for Gaussian and $\textrm{HG}_{10}$ beams are given in Figs. \ref{fig:5}(g) and \ref{fig:5}(h) respectively. The intensity values are normalized with respect to the maximum value of $I$. The simulated FP intensity patterns are in good agreement with their experimental counterparts.

\subsection{Enhancement of SHEL and longitudinal spin density}

As discussed in the previous sections, the spin-Hall signal for $I_-$ and $I_+$ can be obtained through $s_3$. The enhancement of the spin-Hall signal can be further quantified by comparing the $s_3$ distribution for Gaussian and $\textrm{HG}_{10}$ beams. Since the wire is centered at ($x=0$, $y=0$) with its long axis is aligned along $x$ axis and the scattering is prominent in $k_y/k_0$ direction, we consider the $s_3(\kappa_x,\kappa_y)$ distribution along $\kappa_x$, where $\kappa_x = k_x/k_0$ and $\kappa_y = k_y/k_0$. Due to the symmetrical pattern, we consider only one half of the FP i.e., $\kappa_y\geq0$ half. In addition, since the scattering is dominant within the critical angle, we consider the sub-critical region. Thus, similar to Eq. (\ref{eq:4}), quantitative measure of far-field longitudinal spin density distribution ($S_3$) can be obtained by integrating $s_3$ over $\kappa_y$ from $0$ to $1$:
\begin{equation}
S_3(\kappa_x) = \int\displaylimits_{\kappa_y=0}^{1} s_3(\kappa_x,\kappa_y) d\kappa_y.
\label{eq:5}
\end{equation}

\begin{figure}[t]
\includegraphics{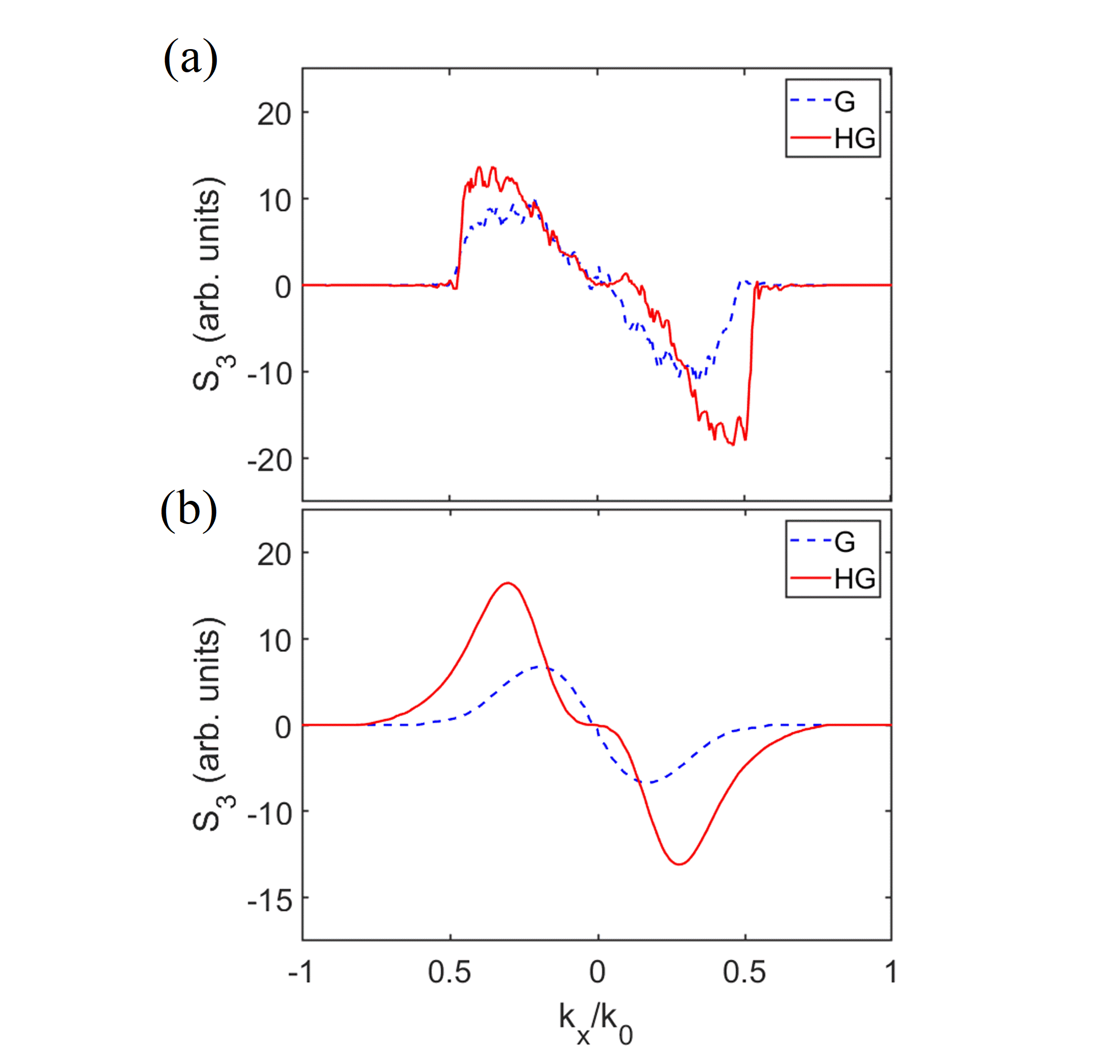}
\caption{\label{fig:6} (color online) Comparative plot of the far-field longitudinal spin density distribution ($S_3(\kappa_x)$) of Gaussian (dashed blue line) and $\textrm{HG}_{10}$ beam (solid red line) extracted from (a) experimentally measured and (b) numerically simulated FP intensity distributions.}
\end{figure}

The  far-field longitudinal spin density distribution, $S_3(\kappa_x)$, extracted from the experimentally measured $s_3$ in Figs. \ref{fig:5}(c) and \ref{fig:5}(d) is plotted in Fig. \ref{fig:6}(a). Similar analysis is also performed for the numerically simulated $s_3$ patterns given in Figs. \ref{fig:5}(g) and \ref{fig:5}(h). Fig. \ref{fig:6}(b) shows the corresponding plot of $S_3(\kappa_x)$. Both the $S_3(\kappa_x)$ extracted from the experimentally measured and numerically calculated $s_3$ exhibit higher longitudinal spin density values and hence higher spin-Hall signal for $\textrm{HG}_{10}$ beam compared to Gaussian beam. The enhancement factor in the spin-Hall signal as well as the longitudinal spin density can be obtained by calculating the ratio of extremum (maximum or minimum) values of $S_3(\kappa_x)$ of $\textrm{HG}_{10}$ to that of Gaussian beam. From Fig. \ref{fig:6}(a) the average enhancement factor value turns out to be, $\eta_{exp}=1.31$. While from the numerically simulated data in Fig. \ref{fig:6}(b), the calculated value of enhancement factor is, $\eta_{sim}= 2.39$. Since focal $s_z \propto s_3$, the enhancement can be attributed to higher $s_z$ of $\textrm{HG}_{10}$ beam with respect to that of Gaussian beam (From Figs. \ref{fig:3}(c) and \ref{fig:3}(d): $\eta_{theory} = 1.29$). The difference between the measured $s_3$ enhancement values $\eta_{exp}$ and $\eta_{sim}$ can be attributed to omission of incident NA region in the experimental data as well as subtle difference of focusing in the numerical simulations with the experimental focusing conditions. 

Hence, our experiments demonstrate SHEL in scattering of linearly polarized focused beam as well as its enhancement for $\textrm{HG}_{10}$ beam compared to Gaussian beam. The enhancement factor is quantified by obtaining the ratio extremum values of spin density distribution $S_3(\kappa_x)$. The complex nature of the optical field at the focus of an objective lens allows generation of longitudinally spinning electric fields which results in such effect.

\section{\label{sec:level5}Conclusion}

In summary, we demonstrate experimental observation of SHEL by analyzing the forward elastic scattered light of focused linearly polarized Gaussian and $\textrm{HG}_{10}$ beams from a single crystalline nanowire. The measure of the spin-Hall signal is obtained by computing far-field longitudinal spin density ($s_3$). Furthermore, by comparing the spin-density distribution ($S_3$) we infer enhancement of spin-Hall signal for $\textrm{HG}_{10}$ beam with respect to Gaussian beam, quantified by the factor $\eta_{exp}=1.31$ from experimental measurements and $\eta_{sim}=2.39$ from numerically simulated results. By studying the focal optical fields, we attribute the observed effect to the generation of longitudinally spinning optical fields due to focusing, hinting the geometrical origin of the effect. Our experiments reveal the very intricate nature of the SOI between a focused optical beam and a nanoscopic object \cite{bliokh2015spin}. In recent times spin-Hall effect of light has gained a lot of traction in research due to great potential for spin assisted photonic devices \cite{ling2017recent,lodahl2017chiral} as well as fundamental study of optical energy flow \cite{Bekshaev_2011, aiello2009transverse}. Additionally, the emerging SOI with a single crystalline AgNW can also pave way for on-chip photonic device applications.

\begin{acknowledgments}
This work was partially funded by Air Force Research Laboratory grant (FA2386-18-1-4118 R\&D 18IOA118), DST Energy Science grant (SR/NM/TP-13/2016) and Swarnajayanti fellowship grant (DST/SJF/PSA-02/2017-18). DP and DKS acknowledge Vandana Sharma, Chetna Taneja, Sunny Tiwari, Shailendra Kumar Chaubey, Surya Narayan Banerjee and Utkarsh Khandelwal for fruitful discussion. 
\end{acknowledgments}

\appendix

\section{\label{sec:ap1}Scattered electric fields at the focal plane}

\begin{figure}[t]
\includegraphics{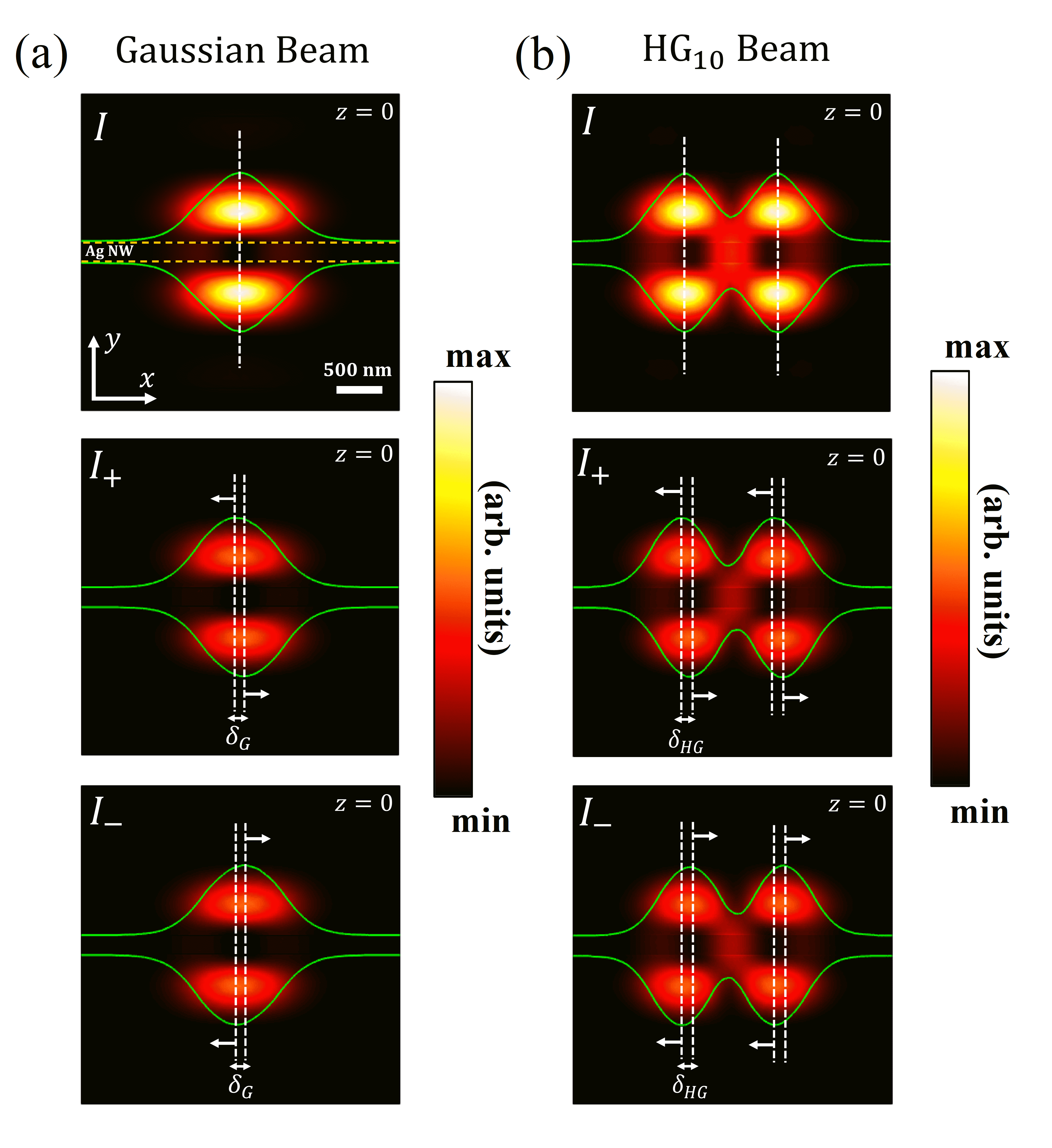}
\caption{\label{fig:ap1} (color online) Numerically simulated scattered normalized electric field intensity at the focal plane ($z=0$) corresponding to (a) Gaussian beam and (b) $\textrm{HG}_{10}$ beam. $I$ represents total scattered electric field, $I_+$ and $I_-$ indicates the LCP and RCP polarized scattered electric fields respectively. The intensity distribution plot (green line) of lobes in $y>0$ and $y<0$ half as a function of $x$ coordinates is super imposed on the figures.}
\end{figure}

Numerically simulated scattered electric field from an AgNW at focal plane ($z=0$ plane) for Gaussian and $\textrm{HG}_{10}$ beams are shown in Fig. \ref{fig:ap1}. The NW positioned at ($x=0$, $y=0$) and along $x$ axis is indicated by the yellow dashed lines. $I$ represents total scattered intensity, $I_+$ and $I_-$ represents the LCP and RCP analyzed components respectively. The intensity distribution plot of the lobe(s) as a function of $x$ coordinates (green lines) in both $y>0$ and $y<0$ halves are superimposed in the Fig. \ref{fig:ap1}.

The scattered intensity profile $I$ reveal that the intensity lobe(s) gets split into two equal halves due to the scattering from the NW. Although there is very less NF accumulation, but with respect to Gaussian, the NF accumulation for $\textrm{HG}_{10}$ beam is more due to higher $E_y$ component (see Fig. \ref{fig:3}(b)). Close inspection of intensity profiles corresponding to $I_+$ and $I_-$ exhibit that with respect to symmetric intensity profile of $I$, the intensity lobe(s) in $y>0$ and $y<0$ regions of $I_+$ and $I_-$ show opposite shifts, as indicated by the white arrows. The spatial shift ($\delta$) along $x$ axis can be quantified by calculating the shift of $I_+$ (or $I_-$) distribution maxima of each of the lobe(s) in $y>0$ half with respect to the corresponding lobe(s) in $y<0$ half. For scattered Gaussian beam the average shift is $\delta_{G} = 64.0$ nm and that for $\textrm{HG}_{10}$ beam is $\delta_{HG}= 92.2$ nm. The enhancement of the positional shift can be obtained as the ratio $\delta_{HG}/\delta_{G} = 1.44$. The average shift calculated here are discussed in more general terms as shift vectors in different systems \cite{PhysRevB.100.201405,PhysRevLett.125.076801}. The average shifts are consequence of breaking of symmetry in the $x$-$y$ plane due to the presence of an extended geometry such as a NW. This spin dependent shift is analogous to SHEL \cite{PhysRevLett.93.083901, hosten2008observation}.

\section{\label{sec:ap2}Wave vector shift in the Fourier plane}

\begin{figure}[t]
\includegraphics{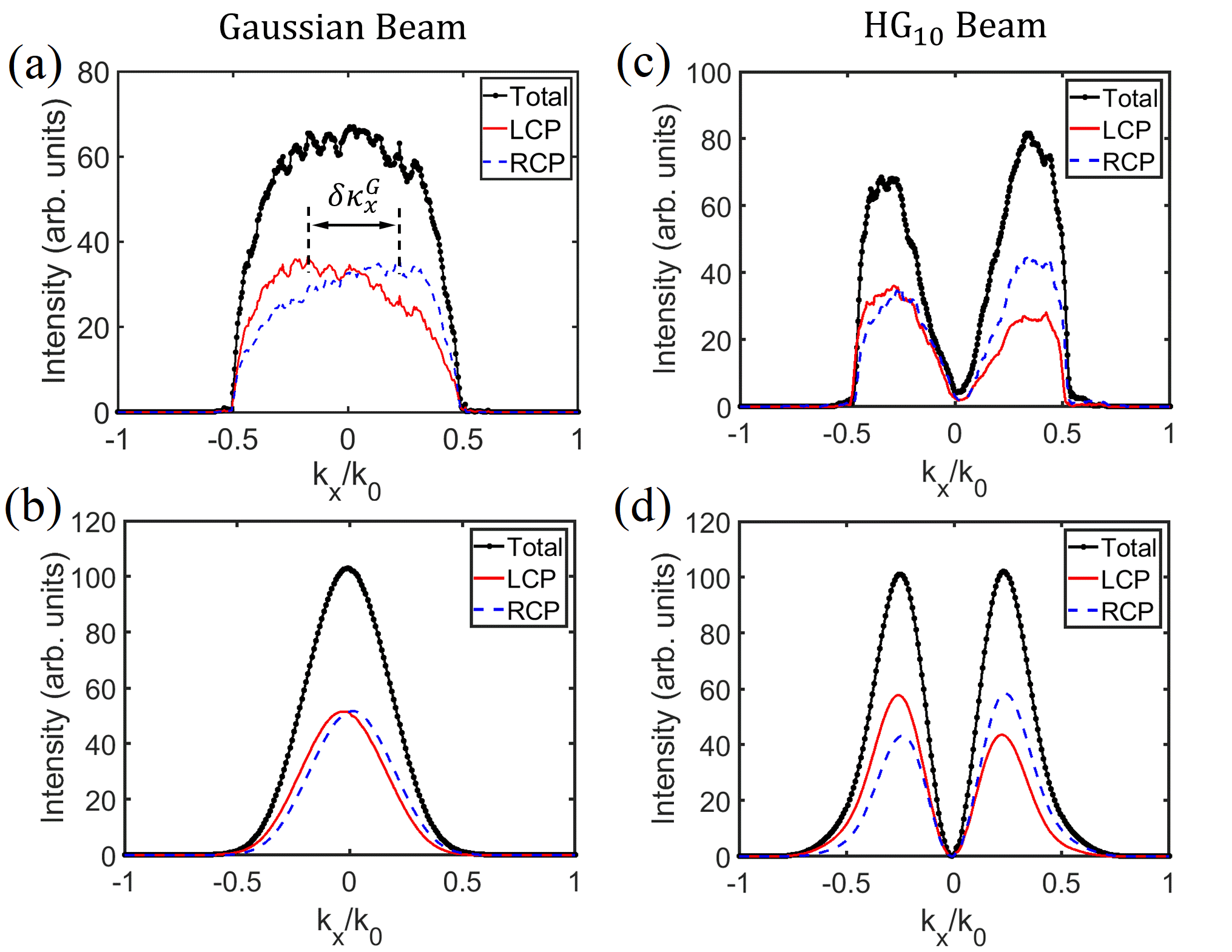}
\caption{\label{fig:ap2} (color online) Intensity distribution plot of $I$ (solid black line) and corresponding $I_+$ (solid red line) and $I_-$ (dashed blue line) components in FP for integrated $\kappa_y=[0,1]$ along $\kappa_x$ for Gaussian beam from (a)experimentally measured data and (b) from numerically simulated FP images. The corresponding distribution plot for $\textrm{HG}_{10}$ beam from experimentally measured data and numerically simulated data is given in (c) and (d) respectively.}
\end{figure}

The FP image of total intensity ($I$) for Gaussian and $\textrm{HG}_{10}$ beam exhibit symmetrical characteristic about $k_x/k_0$ axis, in contrast to anti-symmetrical distribution of $I_+$ and $I_-$. To probe the shift of the intensity lobe(s) of $I_+(I_-)$, we plot the distribution $I_{+(-)}^{tot}(\kappa_x)=\int\displaylimits_{\kappa_y = 0}^1 I_{+(-)}(\kappa_x,\kappa_y)d\kappa_y$ and compare it with the distribution $I^{tot}(\kappa_x)=\int\displaylimits_{\kappa_y = 0}^1 I(\kappa_x,\kappa_y)d\kappa_y$ for Gaussian and $\textrm{HG}_{10}$ beam in Fig. \ref{fig:ap2}. 

For Gaussian beam, we can see from Fig. \ref{fig:ap2}(a) and Fig. \ref{fig:ap2}(b) that the $I^{tot}_+(\kappa_x)$ (solid red line) and $I^{tot}_-(\kappa_x)$ (dashed blue line) analysed distributions exhibit angular shift along $\kappa_x<0$ and $\kappa_x>0$ direction respectively. The measured shift from the experimental data in Fig. \ref{fig:ap2}(a) is $\delta\kappa_{x,expt}^G=0.34$, where as from the numerically simulated FP image the shift measured in Fig. \ref{fig:ap2}(b) is $\delta\kappa_{x,sim}^G=0.05$.

For $\textrm{HG}_{10}$ beam, the shift along $\kappa_x$ is small compared to Gaussian beam as can be seen in Figs. \ref{fig:ap2}(c) and \ref{fig:ap2}(d). The shift along $\kappa_x$, obtained from experimental data in Fig. \ref{fig:ap2}(c) is $\delta\kappa_{x,expt}^{HG}=0.01$ and that obtained from numerically simulated data in Fig. \ref{fig:ap2}(d) is $\delta\kappa_{x,sim}^{HG}=0.01$. For $\textrm{HG}_{10}$ beam, the shift is more along $\kappa_y$ as can be determined from examining the FP intensity distribution corresponding to $I_+$ and $I_-$ in Fig. \ref{fig:5}(f) of manuscript. The presence of nodal line of $\textrm{HG}_{10}$ beam restricts angular shift along $\kappa_x$ axis \cite{sharma2018spin}.

The shift obtained from the experimentally data and numerically simulated data depict a qualitative comparison of the angular shift between Gaussian and $\textrm{HG}_{10}$ beams. The values differ due to difference in exact focusing conditions as well as due to omission of excitation NA region for experimentally measured FP intensity distribution.

\section{\label{sec:ap3}Detailed schematic of the experimental setup}
\begin{figure}[t]
\includegraphics{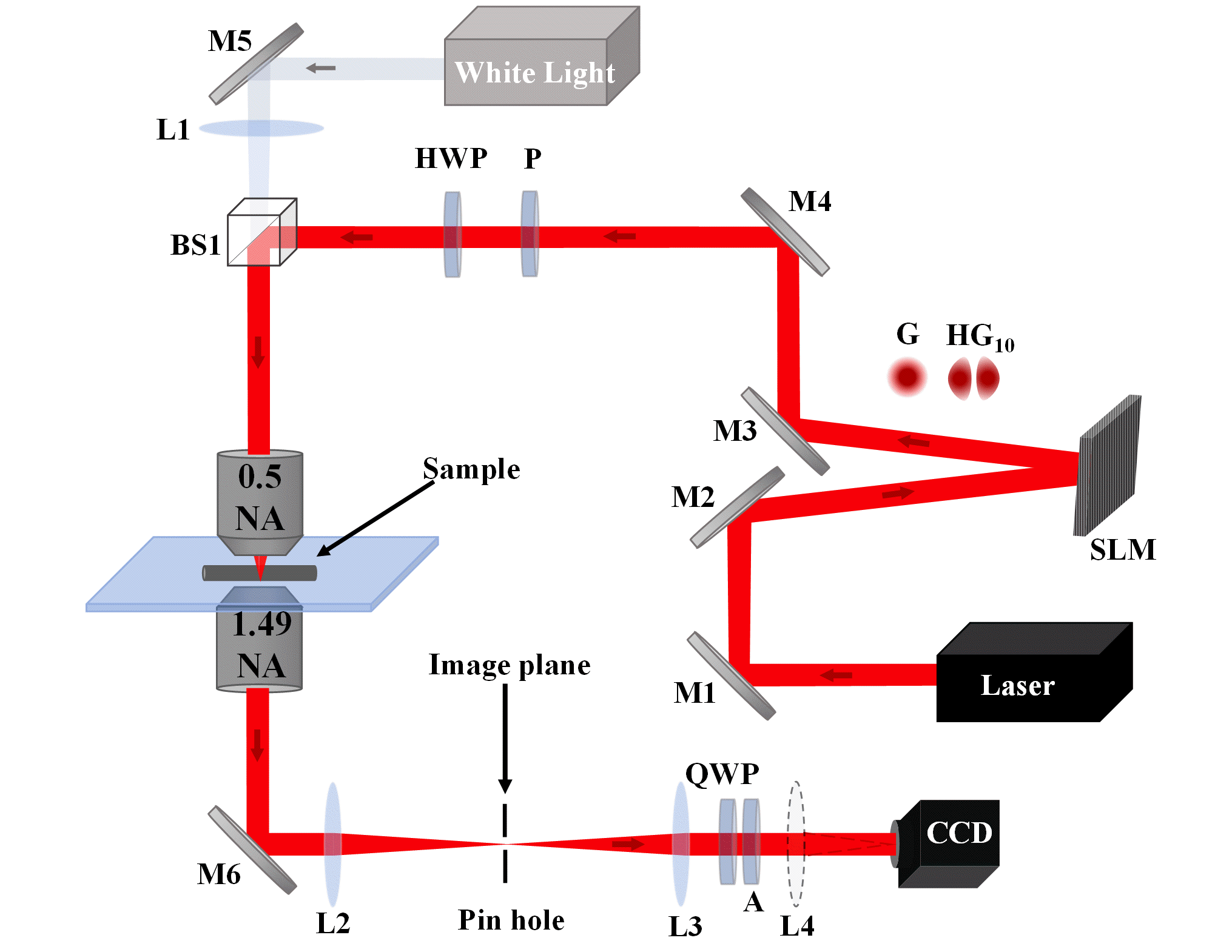}
\caption{\label{fig:ap3} (color online) Detailed schematic of the experimental setup.}
\end{figure}

Detailed schematic of our forward scattering experimental setup is given in Fig. \ref{fig:ap3}. M1-M6 are mirrors, L1-L4 are lenses, HWP: Half-wave plate, QWP: Quarter wave plate, P: Polarizer, A: Analyser, BS1: Beam-splitter. Illumination objective lens ($50\times$, $0.5$ NA) focuses incident Gaussian/$\textrm{HG}_{10}$ beam produced through spatial light modulator (SLM). The scattered light by the sample placed on a piezo-stage is collected using a $100\times$, $1.49$ NA objective lens and back focal plane is imaged at the CCD by projecting it through lenses L2, L3. The collected light is analysed using quarter wave plate and an analyser. A pin hole is placed at the image plane to select the excitation part of the wire and hence to avoid scattering from the other part of the sample. A Flip lens L4 is used to get the real plane image of the sample.


\providecommand{\noopsort}[1]{}\providecommand{\singleletter}[1]{#1}%

\end{document}